\documentstyle[multicol,prl,aps,epsf,floats]{revtex}
\begin{document}
\title{Some novel effects in superconducting nanojunctions}
\author{ Andrei D. Zaikin}
\address{Forshchungszentrum Karlsruhe, Institut f\"ur Nanotechnologie,
76021, Karlsruhe, Germany\\
I.E. Tamm Department of Theoretical Physics, P.N. Lebedev Physics
Institute, 119991 Moscow, Russia}
%\date{\today}
\maketitle

\begin{abstract}
In this paper we address several new developments in the theory of dc
Josephson effect in superconducting weak links. We analyze an interplay
between quantum interference effects and Andreev reflection in $SNS$ 
nanojunctions with insulating barriers and demonstrate that such effects may
qualitatively modify the Josephson current in such structures. We also
investigate an impact of the parity effect on persistent currents in 
superconducting nanorings interrupted by a quantum point contact (QPC). In
the limit of zero temperature and for the odd number of electrons in the ring 
we predict complete suppression of the supercurrent across QPC with 
one conducting mode.  In nanorings
with $SNS$ junctions a $\pi$-state can occur for the odd
number of electrons. Changing this number from even to odd yields
{\it spontaneous} supercurrent in the ground state of such rings without
any externally applied magnetic flux.

\end{abstract}

\begin{multicols}{2}
In 1926 Albert Einstein posed a remarkable question \cite{AE}: ``Of
particular interest is the question whether a link between two
superconductors also turns superconducting''. The answer to
this question was provided by Brian Josephson in 1962 \cite{BJ}.
It was predicted by Josephson that dissipativeless flow of Cooper
pairs between two different superconductors separated by an
insulating barrier is possible provided this supercurrent $I_S$
does not exceed some critical value $I_C$. Furthermore, the
dependence of this current on the phases of macroscopically
coherent wave functions of Cooper pairs was established in a very
simple form \cite{BJ}
\begin{equation}
I_S=I_C\sin \varphi,
\label{dc}
\end{equation}
where $\varphi$ is the difference between the phases of the BCS
order parameters of two superconductors. Eq. (\ref{dc}) represents
the {\it dc Josephson effect}. The answer to the Einstein's
question \cite{AE} turned out to be positive.

What if the total current $I$ flowing through the barrier is larger
than $I_C$? In this case a part of the net current across the barrier
is transferred by normal electrons (quasiparticles) and the rest of
it is carried by Cooper pairs. While the second contribution,
$I_S$, remains dissipativeless and is again described by
Eq. (\ref{dc}), the first -- quasiparticle -- contribution to the
current is dissipative and, hence, causes a non-zero voltage drop $V$
across the insulating barrier. In the presence of this voltage the coherent
phase difference $\varphi$ acquires a time dependence described by
another famous Josephson relation
\begin{equation}
\frac{\partial \varphi}{\partial t}=\frac{2eV}{\hbar}.
\label{ac}
\end{equation}
Combining Eqs. (\ref{dc}) and (\ref{ac}) one immediately arrives
at the conclusion that for any non-zero $V$ the supercurrent $I_S$
changes in time. In the case of time-independent voltages one has
$\varphi =2eVt/\hbar$ and, hence, the Josephson current (\ref{dc})
will oscillate in time with the fundamental frequency proportional
to the voltage $V$. Eq. (\ref{ac}) and related to it oscillations
of the supercurrent represent the essence of the {\it ac Josephson
effect}.

Soon after these Josephson's predictions the microscopic theory
of both dc \cite{AB} and ac \cite{LO,W} was constructed and these
effects have been observed experimentally \cite{YSD}. Huge number
of publications as well as several monographs are devoted to
various aspects of these effects. It turned out that physics
encoded in these phenomena is very rich and important for
understanding of basic properties of superconductivity itself.
More than forty years after its discovery the Josephson effect
still attracts attention of many researchers and keeps providing
us with new interesting physics.

In this paper I will discuss several new phenomena theoretical
understanding of which was achieved only very recently. In the
next section I will very briefly review already well known and
established results which concern dc Josephson effect in various types
of superconducting weak links. Sections II and III are devoted to
possible new effects \cite{GZ,SZ} which emerge and gain importance as 
one decreases the size of a weak link eventually turning it to a
nanostructure with only few conducting channels.  Fabrication of 
such quantum point contacts (QPC) -- unthinkable at the time
of discovery of the Josephson effects -- is now 
becoming a routine procedure. Hence, the new effects discussed
here can be directly observed and investigated in a modern experiment.

\section{Instead of introduction}

Relatively soon after the Josephson's discovery it was understood that
non-dissipative transport of Cooper pairs between two superconductors
is possible not only through a (usually very thin) insulating barrier, but
also in various other situations. One of such situations is realized
in the so-called $SNS$ structures, i.e. if a piece of a normal metal is
placed in-between two superconductors. In contrast to tunnel
junctions, in $SNS$ systems at sufficiently low temperatures appreciable
supercurrent can flow even though  a normal layer can be as thick as
few microns. This is because the wave function of Cooper pairs or, more
precisely, the anomalous Green function, penetrates
into the normal metal from a superconductor at the length $\sim v_F/T$
for ballistic  and $\sim \sqrt{D/T}$ for diffusive metals (here and
below $D=v_Fl/3$ and $l$ are respectively the diffusion coefficient and
the elastic mean free path). Clearly, at temperatures much lower than
the critical temperature $T_c$ of a superconductor this length becomes
large (as compared, e.g., to the superconducting coherence length), and
macroscopic quantum coherence is established between two superconducting
banks separated by a normal metal.

Further studies revealed an interesting
mechanism of Cooper pair transfer in such systems. It turned out that
the supercurrent flow is directly related to another fundamentally important
phenomenon: Andreev reflection \cite{AR}. Suffering Andreev reflections at
both $SN$ interfaces, quasiparticles with energies below the superconducting
gap are effectively ``trapped'' inside the $N$-layer and form a discrete set
of levels \cite{AR}. It was demonstrated \cite{Kulik}
that in the presence of the phase difference $\varphi$ across
the $SNS$ junction these levels acquire a shift proportional to this
phase difference. Thus, on one hand, the position of the quasiparticle energy
levels in such systems can be tuned by passing the supercurrent and, on the
other hand, the magnitude of this supercurrent can be established by
taking the derivative of the quasiparticle energy with respect
to $\varphi$ with subsequent summation over the whole energy spectrum. The microscopic theory \cite{Kulik,Ishii} leads to the following
expression for the current density through clean $SNS$ systems:
\begin{equation}
j=\frac{e^2p_F^2v_F}{6\pi^2d}\varphi , \;\;\; -\pi < \varphi < \pi .
\label{SNST0}
\end{equation}
This expression is valid at $T \to 0$ and for $N$-metal layers with
thickness $d \gg \xi_0 \sim v_F/\Delta$. The most important features
of this result are (i) the strongly non-sinusoidal current-phase relation,
cf. Eqs. (\ref{dc}) and (\ref{SNST0}) and (ii) the linear dependence of 
the current on the gap in the quasiparticle spectrum $\epsilon_{qp} 
\sim v_F/d$ in the direction normal to $NS$ interfaces.

It is interesting that qualitatively both features (i) and (ii) survive
not only for ballistic but also for diffusive $SNS$ junctions even though
in the latter case discrete Andreev levels are washed out due to
elastic scattering of quasiparticles on impurities
in the $N$-metal. It was demonstrated microscopically \cite{Likh,ZZhFNT,us}
that at low temperatures $T \ll D/d^2$ the current-phase relation
in diffusive $SNS$ junctions also deviates from the sinusoidal one
\cite{FN} and the critical Josephson current is again
proportional to the gap in the quasiparticle spectrum, in this case
the Thouless energy $\epsilon_{qp}=D/d^2$. The exact value of the
critical Josephson current in long diffusive $SNS$ junctions can be
established only  numerically. One finds \cite{us}
\begin{equation}
I_c=10.82 \frac{\epsilon_{qp}}{eR_N}
\end{equation}
where $R_N$ is the junction normal state resistance.

The above results -- both for ballistic and diffusive limits -- are valid
for sufficiently long junctions. One can also decrease the thickness
of the normal metal $d$ and gradually crossover to the limit of short
superconducting constrictions. A microscopic description of the dc Josephson
effect of such type of weak links was developed by Kulik and Omel'yanchuk
\cite{KO}. Also in such systems at low temperatures
the current-phase dependence deviates from $\sin \varphi$ and the critical
current $I_c(T\to 0)$ is again proportional to the combination
$\epsilon_{qp}/eR_N$, where now $\epsilon_{qp}=\Delta$. A crossover between
the two limits of long $SNS$ junctions and short superconducting weak links
can also be described microscopically. In the clean case this task can
be trivially accomplished by solving the Eilenberger equations 
\cite{Eil,lar,Albert}, while in the dirty limit one should make use of 
the Usadel equations \cite{Usadel} which can be solved only numerically. 
The latter task has recently been carried out in Ref. \cite{us}.

Let us also note that in all the above considerations
inter-metallic interfaces were assumed to be perfectly
transparent. It is also straightforward to generalize the analysis
in order to account for electron scattering at the insulating
barrier which can be present inside a weak link. For short
superconducting junctions containing an insulating barrier with an
arbitrary energy independent transmission the corresponding
generalization has been worked out by Haberkorn et al. \cite{Hab}.
This analysis yields a general formula for the Josephson current
which matches with the Ambegaokar-Baratoff result \cite{AB} in the weak
tunneling limit and crosses over to the Kulik-Omel'yanchuk's
expression \cite{KO} for clean constrictions at transmissions
approaching unity. It is interesting that the result \cite{KO} for
diffusive constrictions can also be recovered from the formula of
Ref. \cite{Hab} after its slight generalization. In order to do so
one should assume that the transmission is not the same for all
conducting channels but rather obeys the Dorokhov's distribution
formula. Combining this formula with the expression \cite{Hab} and
summing over all conducting channels one arrives at the result
\cite{KO} for diffusive weak links.

One can also investigate transport properties of more complicated layered
structures which contain both normal metal layers and insulating barriers.
For instance, $SNS$ systems with one insulating barrier, such as $SINS$ and
$SNINS$ were analyzed by a number of authors
\cite{ZZh,ZZh2,svi,ZZh1,KuLu,SaMi,SaMi2}. For an extended review 
summarizing various features of dc Josephson effect in
different types of superconducting weak links and further references we refer the reader to Refs. \onlinecite{lam,bel,SaMiZhe}.

Most of the results reviewed above were obtained already long time
ago and are by now well established and well understood. One can
think that considering dc Josephson effect in even more
complicated structures like, for instance, $SNS$ structures with
two or three insulating barriers, may at most yield somewhat more
cumbersome expressions but would not allow to encover any new
physics beyond what has already been understood in simpler
situations. Below we will show that it is not so. Just on the
contrary, in the next section we will demonstrate that qualitatively new 
effects may occur in $SNS$ junctions with more than one insulating barriers,
in particular provided the cross section of such junctions is reduced to be 
comparable to the square of the Fermi wavelength.

\section{Josephson effect and quantum interference of quasiparticles}

In this section we will analyze the dc Josephson effect in $SNS$ systems which
contain several insulating barriers. In this case electrons scattered at
different barriers can interfere inside the junction. We will demonstrate that
such interference may lead to qualitatively new effects and cause
severe modifications of the supercurrent across the junction. We will see that
these modifications can go in both directions, i.e. the Josephson
current can be dramatically decreased by {\it destructive intereference}
of quasiparticles or, on the contrary, increased as a result of their {\it
  constructive intereference}. The first situation is realized for
sufficiently short junctions, while for longer ones the second effect
might become more pronounced.

The phenomenon of quantum interference of quasiparticles is of primary
importance for $SNS$ structures with few conducting channels. The interest
to such structures grew  considerably after several experimental groups
\cite{Kas,Basel1,Basel2} have succeeded in connecting a carbon nanotube to
two superconductors and performing transport measurements in such systems.
More conventional $SNS$ structures with many
conducting channels and several insulating barriers are also of
considerable interest, for instance in relation to possible
applications, see e.g. Ref. \onlinecite{Misha} and further references
therein. We will demonstrate that for such systems
quantum interference effects are also important
provided there exist more than two scatterers inside the junction.

On a theoretical side a significant difficulty is that the powerful
formalism of quasiclassical
energy-integrated Eilenberger Green functions \cite{Eil,lar,Albert,bel}
supplemented by the Zaitsev boundary conditions \cite{zait} cannot be
directly applied to systems containing more than one insulating
barrier. An important ingredient of the derivation\cite{zait} is the
assumption that such barriers are located sufficiently far from each other,
so that {\it interference effects} emerging from electron scattering
can be totally neglected. It is also essential that Zaitsev boundary
conditions do not depend on the scattering phases. Since here we are
just interested in investigating of quantum interference of quasiparticles
we are not in a position to use the quasiclassical Eilenberger
formalism for our purposes. One possibility to circumvent this problem
is to apply the formalism \cite{PZ,GuZ94} within which the presence of an
arbitrary number of barriers in the system can be accounted for by linear
boundary conditions. Another -- even more straightforward -- possibility to 
analyze the dc Josephson effect in structures with several insulating barriers
 is to directly solve the exact Gor'kov equations \cite{AGD}. Here we will
follow the second approach.

The results presented in this section
were obtained in collaboration with A.V. Galaktionov \cite{GZ}. A similar
approach has also been used independently by Brinkman and Golubov \cite{brink}.

\subsection{General formalism}

In what follows we will
assume that our system is uniform along the directions parallel to
the interfaces (coordinates $y$ and $z$).
Performing the Fourier transformation  of the normal
$G$ and anomalous $F^+$ Green function with respect to these coordinates
$$ G_{\omega_n}(\bbox{r}, \bbox{r'})=
\int\frac{d^2 \bbox{k_\parallel}}{(2\pi)^2}
G_{\omega_n} (x,x',\bbox{k_\parallel})e^{i
\bbox{k_\parallel}(\bbox{r_\parallel}- \bbox{r'_\parallel})}
$$
we express the Gor'kov equations in the following standard form
\begin{equation}
{\small \left( \begin{array}{cc} i\omega_n -\hat H & \Delta(x)\\ \Delta^*(x)&
i\omega_n +\hat H_c\end{array}\right)\left( \begin{array}{c} G_{\omega_n}
(x,x',\bbox{k_\parallel})\\F^+_{\omega_n} (x,x',\bbox{k_\parallel})
\end{array}\right)= \left(\begin{array}{c} \delta(x-x')\\ 0\end{array}
\right).} \label{start}
\end{equation}
Here $\omega_n=(2n+1)\pi T$ is the Matsubara frequency, and $\Delta(x)$ is the
superconducting order parameter.The Hamiltonian $\hat H$ in
Eq.(\ref{start}) reads
\begin{equation}
\hat H=-\frac{1}{2m}\frac{\partial^2}{\partial x^2}+
\frac{\bbox{\tilde k^2_\parallel}}{2m}-\epsilon_F +V(x).
\label{H}
\end{equation}
Here $\bbox{\tilde k_\parallel}= \bbox{k_\parallel}-
\frac{e}{c}\bbox{A_\parallel}(x)$, $\epsilon_F$ is Fermi energy,
the term $V(x)$ accounts for the external potentials (including
the boundary potential). The Hamiltonian $\hat H_c$ is obtained
from $\hat H$ (\ref{H}) by inverting the sign of the electron
charge $e$.

As usually, it is convenient to separate fast oscillations of the
Green functions $\propto \exp(\pm ik_x x)$ from the envelope of
these functions changing at much longer scales as compared to the
atomic ones. Then one can construct a particular solution of the
Gor'kov equations (\ref{start}) in the following form
\begin{eqnarray}
\left( \begin{array}{c} G_{\omega_n}
(x,x',\bbox{k_\parallel})\\F^+_{\omega_n}
(x,x',\bbox{k_\parallel})\end{array}\right)=\overline{\varphi}_{+1}(x)
g_1(x') e^{ik_x(x-x')}+ \nonumber\\\overline{\varphi}_{-2}(x)
g_2(x') e^{-ik_x(x-x')}\quad \mbox{if}\: x>x' \label{xbp}
\end{eqnarray}
and
\begin{eqnarray}
\left( \begin{array}{c} G_{\omega_n}
(x,x',\bbox{k_\parallel})\\F^+_{\omega_n}
(x,x',\bbox{k_\parallel})\end{array}\right)=\overline{\varphi}_{-1}(x)
f_1(x') e^{-ik_x(x-x')}+ \nonumber\\\overline{\varphi}_{+2}(x)
f_2(x') e^{ik_x(x-x')}\quad \mbox{if}\: x<x'\label{xmp}.
\end{eqnarray}
These functions satisfy Gor'kov equations at $x\neq x'$. Here
$\overline{\varphi}_+$ are two linearly independent solutions of
the equation
\begin{equation}\left(
\begin{array}{cc} i\omega_n -\hat H^{a}_{\pm} & \Delta(x)\\ \Delta^*(x)&
i\omega_n +\hat H^{a}_{\pm c} \end{array}\right)
\overline{\varphi}_\pm=0. \label{appr}
\end{equation}
The solution $\overline{\varphi}_{+1}$) does not diverge at
$x\rightarrow +\infty$, while $\overline{\varphi}_{+2}$ is
well-behaved at $x\rightarrow -\infty$. Similarly, two linearly
independent solutions $\overline{\varphi}_{-1,2}$ do not diverge
respectively at $x\rightarrow -\infty$ and $x\rightarrow
+\infty$.

In eq. (\ref{appr}) we defined
\begin{equation}
\hat H^{a}_\pm=\mp
iv_x\partial_x-\frac{e}{c}\bbox{A_\parallel}(x)\bbox{v_\parallel} +
\frac{e^2}{2m c^2}\bbox{A_\parallel}^2(x) +\tilde V(x).
\end{equation}
Here  $k_x=mv_x=\sqrt{k_F^2-k_\parallel^2}$, $\tilde V(x)$
represents a slowly varying part of the potential which {\it does
not} include fast variations which may occur at metallic
interfaces. The latter will be accounted for by the boundary
conditions to be considered below.

The functions $f_{1,2}(x)$ and $g_{1,2}(x)$ are determined with
the aid of the continuity condition for the Green functions at
$x=x'$ and the condition resulting from the integration of
$\delta(x-x')$ in eq.(\ref{start}).

A general solution of the Gor'kov equations has the form
\begin{eqnarray}
&\left( \begin{array}{c} G_{\omega_n} (x,x')\\F^+_{\omega_n} (x,x')
\end{array}\right)= \left( \begin{array}{c}
G_{\omega_n} (x,x')\\F^+_{\omega_n} (x,x')\end{array}\right)_{part}+&\nonumber
\\ &[ l_1(x')\overline{\varphi}_{+1}(x) +l_2(x')\overline{\varphi}_{+2}(x)]
e^{ik_x x}+& \label{general} \\ &[l_3(x')\overline{\varphi}_{-1}(x)+
l_4(x')\overline{\varphi}_{-2}(x)]e^{-ik_x x}.&\nonumber
\end{eqnarray}
For systems which consist of several metallic layers the
particular solution is obtained with the aid of the procedure
outlined above provided both coordinates $x$ and $x'$ belong to
the same layer. Should $x$ and $x'$ belong to different layers,
the particular solution is zero because in that case the
$\delta$-function in eq. (\ref{start}) fails. The functions
$l_{1,2,3,4}(x')$ in each layer should be derived from the proper
boundary conditions. These are just the matching conditions for
the wave functions on the left and on the right side of a
potential barrier, respectively $A_1 \exp(ik_{1x}x)+ B_1
\exp(-ik_{1x}x)$ and $A_2 \exp(ik_{2x}x)+ B_2 \exp(-ik_{2x}x)$. is
These conditions  have the standard form (see e.g
\onlinecite{LL}):
\begin{eqnarray}
&& A_2=\alpha A_1+\beta B_1,\: B_2=\beta^*A_1 +\alpha^* B_1,\nonumber
\\
&& |\alpha|^2-|\beta|^2=\frac{k_{1x}}{k_{2x}}. \label{scatt}
\end{eqnarray}
The equations
\begin{equation}
R=\left|\frac{\beta}{\alpha}\right|^2,\quad
D=1-R=\frac{k_{1x}}{k_{2x}|\alpha|^2} \label{scatt2}
\end{equation}
define respectively the reflection and transmission coefficients of
the barrier. Applying these boundary conditions at each insulating
barrier one uniquely determines all the unknown functions in eq.
(\ref{general}) and thereby completes the construction the Green
functions of our problem. For further details we refer the reader
to Ref. \cite{GZ}.

We are now in a position to specify the general expression for the
Josephson current across ballistic $SNS$ junctions which contain
an arbitrary number of insulating barriers. In what follows we
will assume that a thin specularly reflecting insulating barriers
($I$) are situated at both $SN$ interfaces. Additional such
barriers can also be present inside the $N$-metal. Transmissions
of these barriers my take any value from zero to one. We also
assume that electrons propagate ballistically between any two
adjacent barriers and that no electron-electron or electron-phonon
interactions are present in the normal metal. For simplicity we
will restrict our attention to the case of identical
superconducting electrodes with singlet isotropic pairing and
neglect suppression of the superconducting order parameter
$\Delta$ in the electrodes close to the $SN$ interface. The phase
of the order parameter is set to be $-\varphi /2$ ($+\varphi /2$) in the left
(right) electrode. As before, the
thickness of the normal layer will be denoted by $d$.

Employing the standard formula for the current density
\begin{equation}
 J=\frac{ie}{m} T\sum_{\omega_n}\int
\frac{d^2k_\parallel}{(2\pi)^2}\left(\nabla_{x'}- \nabla_x\right)_{x'\to
x}G_{\omega_n} (x,x',\bbox{k_\parallel}). \label{ccuu}
\end{equation}
and making use of the expressions for the Green functions, one
arrives at the following result
\begin{equation}
J=4e T \sum_{\omega_n>0}\int_0^{k_F} \frac{k_xdk_x}{2\pi}
\frac{\sin\varphi}{\cos\varphi +W}, \label{J}
\end{equation}
where the function $W$ depends on the number of insulating
barriers. This function will be specified below for the case of
two and three barriers.

Note that the integral over $k_x$ in eq. (\ref{J}) can be replaced
by a sum over independent conducting channels
\begin{equation}
\frac{{\cal A}}{2\pi}\int_0^{k_F}k_x dk_x(...) \to\sum_m^N (...),
\label{Dm}
\end{equation}
where ${\cal A}$ is the junction cross section. In this case
$D_{1,2}$ and $R_{1,2}$ may also depend on the channel index $m$.

\subsection{$SINI'S$ junctions with few conducting channels}

Let us first consider $SNS$ junctions with two insulating
barriers, one at each $NS$ interface. In this case the function
$W$ in (\ref{J}) takes the form
\begin{eqnarray}
W=\frac{4\sqrt{R_1R_2}}{D_1D_2}\frac{\Omega_n^2}{\Delta^2}\cos\chi
\nonumber
\\+\frac{\Omega_n^2
(1+R_1)(1+R_2)+ \omega_n^2D_1D_2}{D_1D_2\Delta^2}
\cosh\frac{2\omega_nd}{v_x}\label{W}\\
+\frac{2(1-R_1R_2)}{D_1D_2}
\frac{\Omega_n\omega_n}{\Delta^2}\sinh\frac{2\omega_nd}{v_x}.
\nonumber
\end{eqnarray}
Here $\chi=2k_xd+\phi$ is the phase of the product
$\alpha_2^*\beta_2\alpha_1^*\beta_1^*$. Eqs. (\ref{J}), (\ref{W})
provide a general expression for the dc Josephson current in
$SINI'S$ structures valid for arbitrary transmissions $D_1$ and
$D_2$.

Let us first  analyze the above result for the case of one
conducting channel $N=1$. We observe that the first term in
eq. (\ref{W}) contains $\cos(2k_x d+\phi)$ which oscillates
at distances of the order of the Fermi wavelength. Provided at least
one of the barriers is highly transparent and/or
(for sufficiently long junctions $d \gtrsim \xi_0$) the temperature is high
$T \gg v_F/d$ this oscillating term is unimportant and can be
neglected. However, at lower transmissions of both barriers
and for relatively short
junctions $d \lesssim v_F/T$ this term turns out to be of the same order
as the other contributions to $W$ (\ref{W}). In this case
the supercurrent is sensitive to the exact positions of the discrete energy
levels inside the junction which can in turn vary considerably
if $d$ changes at the atomic scales $\sim 1/k_F$. Hence, one can
expect sufficiently strong sample-to-sample fluctuations of the
Josephson current even for junctions with nearly identical
parameters.

Let us first consider the limit of relatively short $SINI'S$
junctions in which case we obtain
\begin{equation}
I=\frac{e\Delta}{2}\frac{{\cal T}\sin \varphi}{{\cal D}} \tanh
\left[\frac{{\cal D}\Delta }{2T}\right], \label{I}
\end{equation}
where we defined
\begin{equation}
{\cal D}(\varphi )=\sqrt{1-{\cal T}\sin^2(\varphi /2)} \label{D1}
\end{equation}
and an effective normal transmission of the junction
\begin{equation}
{\cal T}=\frac{D_1D_2}{1+R_1R_2+2\sqrt{R_1R_2}\cos\chi }.
\label{T}
\end{equation}
Eq. (\ref{I}) has exactly the same functional form as the
result derived by Haberkorn {\it et al.} \cite{Hab}
for $SIS$ junctions with an arbitrary transmission
of the insulating barrier. This result is recovered from our eqs.
(\ref{I}), (\ref{T}) if we assume
e.g. $D_1 \ll D_2$ in which case the total transmission (\ref{T})
reduces to ${\cal T} \simeq D_1$.

As we have already discussed the total transmission ${\cal T}$ and,
hence, the Josephson current fluctuate depending on the exact position
of the bound states inside the junction. The resonant transmission is
achieved for $2k_xd +\phi=\pm\pi$, in which case we get
\begin{equation}
{\cal T}_{\rm res}=\frac{D_1D_2}{(1-\sqrt{R_1R_2})^2}.
\label{Tres}
\end{equation}
This equation demonstrates that for symmetric junctions $D_1=D_2$ at
resonance the Josephson current does not depend on the barrier
transmission at all. In this case ${\cal T}_{\rm res}=1$ and our
result (\ref{I}) coincides with the formula derived by Kulik and
Omel'yanchuk\cite{KO} for ballistic constrictions. In the limit
of low transmissions $D_{1,2} \ll 1$ we recover the standard
Breit-Wigner formula ${\cal T}_{\rm res}=4D_1D_2/(D_1+D_2)^2$
and reproduce the result obtained by Glazman and Matveev\cite{glazman}
for the problem of resonant tunneling through a single Anderson
impurity between two superconductors.

Note that our results (\ref{I}-\ref{T}) also support the conclusion
reached by Beenakker\cite{Ben} that the Josephson current across
sufficiently short junctions has a universal form and depends only on the
total scattering matrix of the weak link which can be evaluated in the
normal state. Although this conclusion is certainly correct in the
limit $d \to 0$, its applicability range depends significantly on the
physical nature of the scattering region. From eqs. (\ref{J}),
(\ref{W}) we observe that the result (\ref{I}), (\ref{D1}) applies
at $d \ll \xi_0$ not very close to the resonance. On the other hand,
at resonance the above result is
valid only under a more stringent condition
$d \ll \xi_0D_{\rm max}$, where we define $D_{\rm max}=$max$(D_1,D_2)$.

Now let us briefly analyze the opposite limit of sufficiently long junctions
$d\gg \xi_0$. Here we will restrict ourselves to the most interesting case
$T=0$. From eqs. (\ref{J}), (\ref{W}) we obtain
\begin{eqnarray}
&& I=\frac{ev_x\sin\varphi}{\pi d
z_1}\left[\frac{\arctan\sqrt{z_2/z_1}}{\sqrt{z_2/z_1}}\right],
\\&& z_{1,2}=\cos^2(\varphi/2)+\frac{1}{D_1 D_2}\left( R_+\pm 2\sqrt{R_1 R_2}
\cos(\chi) \right), \nonumber
\end{eqnarray}
where $R_+=R_1+R_2$. For a fully transparent channel
$D_1=D_2=1$ the above expression
reduces to the well known Ishii-Kulik result\cite{Ishii,Kulik}
\begin{equation}
I=\frac{ev_x\varphi}{\pi d},\quad -\pi<\varphi<\pi,
\end{equation}
whereas if one transmission is small $D_1 \ll 1$ and $D_2 \approx 1$
we reproduce the result\cite{ZZh}
\begin{equation}
I=\frac{ev_xD_1\sin\varphi}{2d}. \label{Z}
\end{equation}
Provided the transmissions of both $NS$-interfaces are low $D_{1,2}\ll
1$ we obtain in the off-resonant region
\begin{equation}
I=\frac{ev_x}{4\pi d}D_1D_2\sin\varphi\Upsilon[\chi ],
\end{equation}
where $\Upsilon[\chi ]$ is a $2\pi$-periodic function defined as
\begin{equation}
\Upsilon[\chi ]=\frac{\chi }{\sin\chi },\quad  -\pi<\chi <\pi.
\end{equation}
In the vicinity of the resonance $||\chi |-\pi
 |\lesssim D_{\rm max}$ the above result does not hold
anymore. Exactly at resonance $\chi =\pm \pi$ we get
\begin{equation}
I=\frac{ev_x\sqrt{D_1 D_2} \sin\varphi}{4d
\left\{\cos^2\frac{\varphi}{2}+\frac{1}{4}\left(
\sqrt{\frac{D_1}{D_2}}-\sqrt{\frac{D_2}{D_1}}\right)^2\right\}^{1/2}}.
\end{equation}
For a symmetric junction $D_{1,2}=D$ this formula yields
\begin{equation}
I=\frac{ev_xD \sin(\varphi/2)}{2d},\quad  -\pi<\varphi<\pi,
\end{equation}
while in a strongly asymmetric case $D_1 \ll D_2$ we again arrive
at the expression (\ref{Z}). This implies that at resonance the
barrier with higher transmission $D_2$ becomes effectively transparent
even if $D_2 \ll 1$. We conclude that for $D_{1,2} \ll 1$ the maximum
Josephson current is proportional to the product of transmissions
$D_1D_2$ off resonance, whereas exactly at resonance it is
proportional to the lowest of two transmissions $D_1$ or $D_2$.

We observe that both for short and long $SINI'S$ junctions interference
effects may enhance the Josephson effect or partially suppress it
depending on the exact positions of the bound states inside the junction. We
also note that in order to evaluate the supercurrent across $SINI'S$ junctions
it is in general {\it not} sufficient to derive the transmission probability
for the corresponding $NINI'N$ structure. Although the normal
transmission of the above structure is given by eq. (\ref{T}) for {\it all}
values of $d$, the correct expression for the Josephson current can be
recovered by combining eq. (\ref{T}) with the results\cite{Hab,Ben}
in the limit of short junctions $d \ll D\xi_0$ only. In this case one can
neglect suppression of the anomalous Green functions inside the normal layer
and, hence, the information about the normal transmission turns out to be
sufficient. On the contrary, for longer junctions the decay of Cooper pair
amplitudes inside the $N$-layer cannot be anymore disregarded. In this case
the supercurrent will deviate from the form (\ref{I}) even though the normal
transmission of the junction (\ref{T}) will remain unchanged. This deviation
becomes particularly pronounced for long junctions, i.e. for $d \gg \xi_0$ out
of resonance and for $d \gg D\xi_0$ at resonance.

Generalization of the above results to the case of an
arbitrary number of independent conducting channels $N >1$
is trivial: The supercurrent is simply given by
the sum of the contributions from all the channels.
These contributions are in general not
equal because the phase factors $\chi=2k_xd+\phi$ change randomly
for different channels. Hence, mesoscopic fluctuations of
the supercurrent should become smaller with increasing number of
channels and eventually disappear in the limit of large $N$.

In the latter limit the Josephson current is obtained by averaging
over all values of the phase $\chi$. This limit was already studied
in details\cite{GZ,brink} and will not be considered here.
We will only point out that -- as it was demonstrated in Ref. \onlinecite{GZ}
-- in the limit $N \to \infty$ interference effects are effectively
averaged out and exactly the same result can be reproduced by means
of the Eilenberger formalism supplemented by Zaitsev boundary conditions.
We also worthwhile to emphasize that the latter statement applies
only to the junctions with two insulating barriers. Below we will
show that for system with more than two barriers quasiparicle
interference effects turn out to be even more significant, and the
correct result for the current cannot be recovered with the aid of
Zaitsev boundary conditions even in the limit $N \to \infty$.

\subsection{Josephson current in $\bbox{SINI'NI''S}$ junctions}

Let us now turn to  $SNS$ structures with three insulating barriers. As
before, two of them are located at $SN$ interfaces, and the third
barrier is inside the $N$-layer at a distance $d_1$ and $d_2$
respectively from the left and right $SN$ interfaces.
The transmission and reflection coefficients of this intermediate
barrier are denoted as $D_0$ and $R_0=1-D_0$, whereas the left and
the right barriers are characterized respectively by $D_1=1-R_1$ and
$D_2=1-R_2$.

The supercurrent is calculated along the same lines as it was done
for the case of two barriers. The final result is again expressed
by eq. (\ref{J}), where the function $W$ is now defined by a
substantially more cumbersome expression than one for the two barriers case.
This expression was evaluated in Ref. \onlinecite{GZ} and will not
be presented here. We will go over to the final results.

\subsubsection{One channel limit}

Let us first discuss
the case of one conducting channel. In the
limit of short junctions $d\ll \xi_0D_{\rm max}$ we again
reproduce the result (\ref{I}) where the total effective
transmission of the normal structure with three barriers takes the
form
\begin{equation}
{\cal T}=\frac{2t_1t_0t_2}{1+t_1t_0t_2+{\cal
C}(\varphi_{1,2},t_{0,1,2})},
\end{equation}
where
$$
{\cal C}=\cos\chi_1\sqrt{(1-t_0^2)(1-t_1^2)}
+\cos\chi_2\sqrt{(1-t_0^2)(1-t_2^2)}
$$
\begin{equation}
+(\cos\chi_1\cos\chi_2-t_0\sin\chi_1\sin\chi_2)
\sqrt{(1-t_1^2)(1-t_2^2)}.
\end{equation}
Here we define $t_{0,1,2}=D_{0,1,2}/(1+R_{0,1,2})$ and
$\chi_{1,2}=2k_xd_{1,2}+\phi_{1,2}$. For later purposes let us
also perform averaging of this transmission over the phases
$\chi_{1,2}$. We obtain
\begin{equation}
\langle {\cal T}\rangle=\frac{2 t_1t_0 t_2}{\sqrt{2  t_1t_0 t_2
+t_1^2t_0^2+t_1^2t_2^2+t_0^2t_2^2-t_1^2t_0^2t_2^2}}.
\label{3t}
\end{equation}
In particular, in the case of similar barriers with small
transparencies $D_{0,1,2}\approx D \ll 1$ the average normal
transmission of our structure is  $\langle {\cal T}\rangle \sim
D^{3/2}$. Suppression of the average transmission below the value
$\sim D$ is a result of destructive interference and indicates the
tendency of the system towards localization.

Let us now proceed to the limit of a long junction $d_{1,2}\gg
\xi_0$ and $T=0$. In the off-resonant region for $d_1=d_2$ we find
\begin{equation}
I=\frac{ev_x D_1D_0D_2 \sin\varphi}{8\pi d_1}
\frac{\Upsilon[\chi_1]-\Upsilon[\chi_2]}{\cos\chi_2-\cos\chi_1}.
\end{equation}
This expression diverges at resonance (i.e. at $\chi_1\simeq \pi$
or $\chi_2 \simeq \pi$) where it becomes inapplicable. In the
resonant region $\chi_2 \simeq \pi$ we obtain
\begin{equation}
I=\frac{ev_x \sqrt{D_1D_0 D_2}\sin\varphi}{4d
\sqrt{2(1+\cos\chi_1)({\cal T}^{-1}-\sin^2(\varphi/2))}}.
\end{equation}

\subsubsection{Many channel junctions}

As it was already discussed, in the many channel limit it is
appropriate to average the current over the scattering phases.
Practically in any realistic physical realization
the widths $d_1$ and $d_2$ fluctuate independently on the
atomic scale. In this case averaging over $\chi_1$ and $\chi_2$
should also be performed independently. If $d_1$ and $d_2$ do not
change on the atomic scale but are incommensurate, independent
averaging over the two phases is to be performed as well. Independent
averaging cannot be fulfilled only in  (physically irrelevant) case of
strictly commensurate $d_1$ and $d_2$ which will not be considered
below.

Technically independent averaging over the scattering phases
$\chi_1=x$ and $\chi_2=\lambda x$ amounts to evaluating the
integral of the expression $1/[t+\cos x\cos(\lambda x)]$ from
$x=0$ to some large value $x=L$. At $\lambda=1$ the result of this
integration is $L/\sqrt{t(1+t)}$. However, if $\lambda$ is
irrational, the integral approaches the value $2LK(1/t^2)/\pi t$,
where $K(h)=F(\pi/2,h)$ is the complete elliptic integral.

Let us assume that the transparencies of all three interfaces are
small as compared to one. After averaging over the two scattering
phases we arrive at the final expression for the current
\begin{equation}
J=\frac{ek_F^2}{\pi^2}D_{\rm eff} \sin\varphi T
\sum_{\omega_n>0}\frac{\Delta ^2}{\Omega_n^2}K\left[
\frac{\Delta^2\sin^2(\varphi/2)}{ \Omega_n^2}\right], \label{kll}
\end{equation}
where we define the effective transmission
\begin{equation}
D_{\rm eff}=\int_0^1 \mu d\mu \sqrt{D_0D_1D_2}.
\label{Deff}
\end{equation}
Hence, for similar barriers we obtain the dependence $J \propto
D^{3/2}$ rather than $J \propto D$ (as it would be the case for independent
barriers). The latter dependence would follow from the calculation
based on Zaitsev boundary conditions for the Eilenberger propagators.
We observe, therefore, that quantum interference effects {\it
decrease} the Josephson current in systems with three insulating
barriers. This is essentially quantum effect which cannot be recovered
from Zaitsev boundary conditions even in the multichannel limit.
This effect has exactly the same origin as a quantum suppression of
the average normal transmission $\langle {\cal T} \rangle$ due to
localization effects. Further limiting expressions for short junctions
can be directly recovered from Eq. (\ref{3t}).

We also note that the current-phase relation (\ref{kll}) deviates
from a pure sinusoidal dependence even though all three
transmissions are small $D_{0,1,2} \ll 1$.  At $T=0$ the critical
Josephson current is reached at $\varphi \simeq 1.7$ which is
slightly higher than $\pi/2$. Although this deviation is
quantitatively not very significant, it is nevertheless important
as yet one more indication of quantum interference of electrons
inside the junction.

Finally, let us turn to the limit of long junctions
$d_{1,2}\gg\xi_0$. We again restrict ourselves to the case of low
transparent interfaces. At high temperatures $T\gg v_F/2\pi
d_{1,2}$  we get $ J\propto
D_0D_1D_2e^{-\frac{d}{\xi(T)}}$, where $d=d_1+d_2$ and
$\xi(T)=v_F/(2\pi T)$. In this case the anomalous
 Green function strongly decays deep in the normal layer. Hence,
interference effects are not important and the interfaces can be
considered as independent from each other. In the opposite limit
$T\ll Dv_F/d$, however, interference
effects become important, and the current becomes proportional to
$D^{5/2}$ rather than to $D^3$. Explicitly, at $T \to 0$ with the
logarithmic accuracy we get
\begin{equation}
J=\frac{e k_F^2 v_F\sin\varphi}{16\pi^2\sqrt{d_1d_2}}\int_0^1d\mu
\mu^2 D_1D_2\sqrt{D_0}\ln D_0^{-1}. \label{inc}
\end{equation}
We see that, in contrast to short junctions, in the limit of thick
normal layers interference effects {\it increase} the Josephson
current as compared to the case of independent barriers. The
result (\ref{inc}), as well as one of Eqs. (\ref{kll})
(\ref{Deff}) cannot be obtained from the Eilenberger approach
supplemented by Zaitsev boundary conditions.

\subsection{Some conclusions}

By directly solving the Gor'kov equations we evaluated the dc
Josephson current in $SNS$ junctions containing two and three insulating
barriers with arbitrary transmissions, respectively $SINI'S$ and
$SINI'NI''S$ junctions. Our results can be directly applied both to the
junctions with few conducting channels (such as, e.g., superconductor-carbon
nanotube-superconductor junctions \cite{Kas,Basel1,Basel2}) and to
more conventional $SNS$ structures in the many channel limit. We have
demonstrated that an interplay between the proximity effect and
quantum interference of quasiparticles may play a crucial role
in such systems causing strong modifications of the Josephson current.

For the system with two barriers and few
conducting channels we found
strong fluctuations of the Josephson critical current
depending on the exact position of the resonant level inside
the junction. For short junctions $d \ll \xi_0D$ at
resonance the Josephson current does not depend on the barrier
transmission $D$ and is given by the standard Kulik-Omel'yanchuk
formula \cite{KO} derived for ballistic weak links. In the limit
of long $SNS$ junctions $d \gg \xi_0$ resonant effects may also
lead to strong enhancement of the supercurrent, in this
case at $T \to 0$ and at resonance the Josephson current is proportional to $D$
and not to $D^2$ as it would be in the absence of interference
effects.

While the above
results for few conducting channels cannot be obtained by means
of the approach employing Zaitsev boundary conditions, in the many
channel limit and for junctions with two barriers the latter
approach {\it does} allow to recover correct results. This is
because the contributions sensitive to the scattering phase
are effectively averaged out during summation over conducting
channels.

Quantum interference effects turn out to be even more
important in the proximity systems which contain three insulating
barriers. In this case the quasiclassical approach based on Zaitsev
boundary conditions fails even in the limit of many conducting
channels. In that limit the Josephson current is {\it decreased}
for short junctions ($J \propto D^{3/2}$) as compared to the
case of independent barriers ($J \propto D$). This effect is caused by
destructive interference of electrons reflected from different
barriers and indicates the tendency of the system towards
localization. In contrast, for long $SNS$ junctions with three
barriers an interplay between quantum interference and proximity
effect leads to enhancement of the Josephson current at $T \to 0$:
We obtained the dependence $J \propto D^{5/2}$ instead of $J \propto
D^3$ for independent barriers.

\section{Parity affected Josephson current}

Let us now turn to a different issue which -- to the best of our
knowledge -- was not yet attracted attention in the literature.
Namely, we will discuss an interplay between the parity effect and
the dc Josephson current in superconducting weak links. The results presented
in this section have been obtained in collaboration with S.V. Sharov \cite{SZ}.

It is well known that thermodynamic properties of isolated
superconducting systems are sensitive to the parity of the total
number of electrons \cite{AN92,Tuo93} even though this number
${\cal N}$ is macroscopically large. This parity effect is a
direct consequence of the fundamental property of a
superconducting ground state described by the condensate of Cooper
pairs. The number of electrons forming this condensate is
necessarily even, hence, for odd ${\cal N}$ at least one electron
always remains unpaired having an extra energy equal to the
superconducting energy gap $\Delta$. At sufficiently low
temperatures a clear difference between the superconducting states
with even and odd ${\cal N}$ was demonstrated experimentally
\cite{Tuo93,Laf93}.

Can the supercurrent be affected by this parity effect? At the
first sight the answer to this question should be negative because
of the fundamental uncertainty relation $\delta {\cal N} \delta
\varphi \gtrsim 1$. Should the electron number ${\cal N}$ be
fixed, fluctuations of the superconducting phase $\varphi$ become
large disrupting the supercurrent in the system. On the other
hand, suppressing fluctuations of the phase $\varphi$ will destroy
the parity effect because of large fluctuations of ${\cal N}$.

Despite that, below we will demonstrate that in certain
superconducting structures the parity effect can
coexist with the non-vanishing supercurrent. Consider a
superconducting system which can support circular persistent
currents (PC). For An example is provided by an isolated superconducting
ring pierced by the magnetic flux $\Phi$ in which case 
circulating PC is induced in the ring.
In accordance with the number-phase uncertainty relation the
global superconducting phase of the ring fluctuates strongly in
this case, however these fluctuations are decoupled from the
supercurrent and therefore can be integrated out without any
influence on the latter. In what follows we will show that the
parity effect may substantially modify PC in superconducting
nanorings, in particular for odd number of electrons.

\subsection{Parity projection formalism}

In order to systematically investigate the influence of the
electron parity number on persistent currents in superconducting
nanorings we will employ the well known parity projection
formalism \cite{JSA94,GZ94,AN94}. Recapitulating the key points of
this approach we will closely follow Ref. \onlinecite{GZ94}.

The grand canonical partition function ${\cal Z}(T,\mu )= {\rm Tr
} e^{-\beta({\cal H}-\mu {\cal N})}$ is linked to the canonical
one $Z(T,{\cal N})$ by means of the following equation
\begin{equation}
\label{6}{\cal Z}(T,\mu )=\sum\limits_{{\cal N}=0}^\infty Z(T,{\cal N})\exp \biggl({\frac{%
\mu {\cal N}}T}\biggr).
\end{equation}
Here and below ${\cal H}$ is the system Hamiltonian,  ${\cal N}$
is the total number of electrons and $\beta \equiv 1/T$. Inverting
this relation and defining the canonical partition functions
$Z_{e}$ and $Z_{o}$ respectively for even (${\cal N}\equiv {\cal
N}_{e}$) and odd (${\cal N} \equiv {\cal N}_{o}$) ensembles, one
gets
\begin{equation}
\label{8}Z_{e/o}(T)={\frac 1{2\pi }}\int\limits_{-\pi }^\pi due^{-i{\cal N}_{e/o}u}%
{\cal Z}_{e/o}(T,iTu), \label{invrel}
\end{equation}
where
$$
{\cal Z}_{e/o}(T,\mu )={1\over 2} {\rm Tr }\left\{\big[1\pm
(-1)^{\cal N}\big] e^{-\beta({\cal H}-\mu {\cal N})}\right\}
$$
\begin{equation}
\label{11}={\frac 12}\bigl({\cal Z}(T,\mu )\pm {\cal Z}(T,\mu +i\pi T)\bigr)%
\label{pm}
\end{equation}
are the parity projected grand canonical partition functions. For
${\cal N} \gg 1$ it is sufficient to evaluate the integral in
(\ref{invrel}) within the saddle point approximation which yields
\begin{equation}
\label{14}Z_{e/o}(T)\sim e^{-\beta (\Omega _{e/o}-\mu _{e/o}{\cal
N}_{e/o})},
\end{equation}
where ${\Omega }_{e/o} =-T\ln {\cal Z}_{e/o}(T,\mu )$ are the
parity projected thermodynamic potentials. They can be presented
in the form
\begin{eqnarray}
{\Omega }_{e/o}= {\Omega }_{f} - T \ln \bigg[ \frac{1}{2} \Big( 1
\pm e^{- \beta ({\Omega_{b}} - {\Omega_{f}})} \Big) \bigg],
\label{Ome/o}
\end{eqnarray}
where
\begin{equation}
{\Omega}_{f/b}= - T \ln \bigg[ {\rm Tr }\left\{(\pm1)^{\cal N}
e^{-\beta({\cal H}-\mu {\cal N})}\right\} \bigg].
%\label{3}
\end{equation}
Chemical potentials $\mu _{e/o}$ are defined by the saddle point
condition $N_{e/o}=-\partial \Omega _{e/o}(T,\mu_{e/o})/\partial
\mu_{e/o}$.

The main advantage of the above analysis is that it allows to
express the canonical partition functions and thermodynamical
potentials in terms of the parity projected grand canonical ones
thereby enormously simplifying the whole calculation. We further
note that $\Omega_{f}$ is just the standard grand canonical
thermodynamic potential and $\Omega_{b}$ represents the
corresponding potential linked to the partition function ${\cal
Z}(T,\mu +i\pi T)$. It is easy to see \cite{GZ94} that in order to
recover this function one can evaluate the true grand canonical
partition function ${\cal Z}(T,\mu )$, express the result as a sum
over the Fermi Matsubara frequencies $\omega_f =2\pi T(m+1/2)$ and
then substitute the Bose Matsubara frequencies  $\omega_b =2\pi
Tm$ instead of the Fermi ones. This procedure will automatically
yield the correct expression for ${\cal Z}(T,\mu +i\pi T)$ and,
hence, for $\Omega_{b}$.

Having found the thermodynamic potentials for the even and odd
ensembles one can easily determine the equilibrium current $I$.
Here we will be interested in describing the currents flowing in
isolated superconducting rings pierced by the external magnetic flux
$\Phi_x$. Then in the case of even/odd total number of electrons one
obtains
\begin{equation}
I_{e/o}=I_{f}\pm \frac{I_{b}-I_{f}}{ e^{\beta ({\Omega_{b}} -
{\Omega_{f}})} \pm 1}, \label{Ie/o}
\end{equation}
where the upper/lower sign corresponds to the even/odd ensemble
and we have defined
$$
I_{e/o}= -c \left( \frac{\partial \Omega_{e/o}}{\partial \Phi_x}
\right) _{\mu(\Phi_x)},\;\;\; I_{f/b}= -c \left( \frac{\partial
\Omega_{f/b}}{\partial \Phi_x} \right) _{\mu(\Phi_x)}.
$$

\subsection{Parity effect in nanorings and blocking of the supercurrent}

Let us now make use of the above general expressions and
investigate the influence of the parity effect on PC in
superconducting nanorings with quantum point contacts (QPC).
Before turning to concrete calculations we shall specify the model
for our system. We shall consider mesoscopic superconducting rings
with cross section $s$ and perimeter $L=2\pi R$. The rings will be
assumed sufficiently thin, i.e. $\sqrt{s} \ll \lambda_L$, where
$\lambda_L$ is the London penetration length. Superconductivity
will be described within the (parity projected) mean field BCS
theory. At sufficiently low temperatures this description is
justified provided quantum phase slips (QPS)
\cite{QPSth,QPSexp,MLG} in nanorings can be neglected. This
requirement in turn implies that the ring cross section should be
sufficiently large. With the aid of the results \cite{QPSth} one
concludes that the QPS tunneling amplitude remains exponentially
small provided the condition $s \gg \lambda_F^2 \sqrt{\xi_0/l}$ is
satisfied. Here $\lambda_F$ is the Fermi wavelength, $\xi_0 \sim
v_F/\Delta$ is the coherence length and $l$ is the electron
elastic mean free path assumed to be shorter than $\xi_0$. For
generic systems QPS effects can usually be neglected provided the
transversal size of the wire/ring $\sqrt{s}$ exceeds $\sim 10$ nm.
Hence, the total number of conducting channels in the ring $N_{\rm
r} \sim s/\lambda_F^2$ should inevitably be large $ N_{\rm r} \gg
1$. In addition, the ring perimeter $L$ should not be too large,
so that one could disregard the QPS-induced reduction of the PC
amplitude \cite{MLG}. Finally, we will neglect the difference
between the mean field values of the BCS order parameter for the
even and odd ensembles \cite{JSA94,GZ94}. This is legitimate
provided the volume of a superconducting ring is large enough,
${\cal V}=Ls \gg 1/\nu\Delta$, where $\nu$ is the density of
states at the Fermi level and $\Delta$ is the BCS order parameter
for a bulk superconductor at $T=0$. All these requirements can
easily be met in a modern experiment.

The task at hand is now to evaluate the thermodynamic potentials
$\Omega_{f/b}$. Within the mean field treatment these quantities
can be expressed in terms of the excitation energies
$\varepsilon_{k}$ and the superconducting order parameter
$\Delta(\mbox{\boldmath$r$})$. One finds \cite{GZ94}
\begin{eqnarray}
\Omega_{f}=  \tilde \Omega &-&2T \sum_{k} \ln\left(2 \cosh
\frac{\varepsilon_{k}}{2T} \right),
\label{Omf}\\
\Omega_{b}= \tilde \Omega &-&2T \sum_{k} \ln\left(2 \sinh
\frac{\varepsilon_{k}}{2T} \right), \label{Omb}
\end{eqnarray}
where $\tilde \Omega  = \int d^3 \mbox{\boldmath$r$}
|\Delta(\mbox{\boldmath$r$})|^2 / g +{\rm Tr}\{\hat \xi\}, $ $g$
is the BCS coupling constant and  $\hat \xi$ is the
single-particle energy operator:
\begin{equation}
\hat \xi=
\frac{1}{2m}{\left(-i\hbar\frac{\partial}{\partial\mbox{\boldmath$r$}}-
\frac{e}{c}\mbox{\boldmath$A$}(\mbox{\boldmath$r$})\right)}^2
+U(\mbox{\boldmath$r$})-\mu ,
\end{equation}
\mbox{\boldmath$A$}(\mbox{\boldmath$r$}) is the vector potential
and $U(\mbox{\boldmath$r$})$ describes the potential profile due
to disorder and interfaces.

The excitation spectrum $\varepsilon_{k}$ has the form
\begin{equation}
\varepsilon_{k}=\varepsilon(\mbox{\boldmath$p$})=\mbox{\boldmath$p$}
\mbox{\boldmath$v$}_{S}+ \sqrt{\xi^{2}+ \Delta^{2}},
\label{exc}
\end{equation}
where $\mbox{\boldmath$p$}$ is a quasiparticle momentum, $\xi
=(p^{2}-\tilde {\mu})/2m$, and $\tilde
{\mu}=\mu(\Phi_x)-{m{v_{S}}^2}/2$. The superconducting velocity
vector $\mbox{\boldmath$v$}_{S}$ is oriented in the direction
along the ring and is defined by the well known expression
\begin{equation}
v_{S}= \frac{\hbar}{2mR}{\rm
min}_n\left(n-\frac{\Phi_x}{\Phi_{0}}\right). \label{st}
\end{equation} This expression as well as the excitation
spectrum (\ref{exc}) are the periodic functions of the flux $\Phi_x$
with the period equal to the superconducting flux quantum
$\Phi_{0}=hc/2e$.

Consider the most interesting case $T \to 0$. Making use of the
above expressions one easily finds
\begin{equation}
I_{e}=ev_{S}\varrho_{e} s,\;\; \mu_{e}=\mu_(\varrho_e)+mv^2_{S}/2
\label{evenI}
\end{equation}
for the even ensemble and
\begin{eqnarray}
I_{o}=ev_{S}\varrho_{o}s-e\frac{v_{F}}{L}\mbox{sgn}(v_{S}), \;\;\;
\varrho_{o}=\varrho +\frac{1}{\cal V} \frac{|v_{S}|}{v_{\mu}}.
\label{resI}
\end{eqnarray}
for the odd one. Here $\varrho_{e/o}={\cal N}_{e/o}/{\cal V}$ are
the electron densities for the even and odd ensembles, $\varrho$
is the grand canonical electron density at $T=0$,
$v_{\mu}=\sqrt{2\mu /m}$ and $v_{S}$ is assumed to be small as
compared to the critical velocity $v_{C}= \Delta/p_{F}$. We also
note that the second Eq. (\ref{resI}) is an implicit equation for
the chemical potential $\mu_{o}$.

Eq. (\ref{evenI}) -- being combined with (\ref{st}) -- coincides with that
obtained for the grand canonical ensemble. In particular, the
current $I_{e}$ represents the well known ``saw tooth'' dependence
on magnetic flux. In contrast, for odd ensembles there exists an
additional flux-dependent contribution to PC (\ref{resI}) which
cannot be viewed just as a renormalization of $\varrho_{o}$.

Unfortunately this parity effect is rather small in multichannel
rings \cite{FN}. Estimating the leading contribution to $I_{e/o}$
as $I\sim ev_{F}N_{\rm r}/L$, we find
$$
(I_{e}-I_{o})/I \sim 1/N_{\rm r} \ll 1.
$$

The results (\ref{evenI})-(\ref{resI}) hold as long as $T \ll
\hbar v_{F}/L$. At higher temperatures the parity effect gets even
smaller and eventually disappears at temperatures exceeding the
parameter \cite{Tuo93} $T^*\approx \Delta/\ln (\nu {\cal
V}\sqrt{\Delta T^*})$. The corresponding expressions are readily
obtained within our formalism, but we will not consider them here.

Rather we turn to a somewhat different system -- a superconducting 
ring interrupted by QPC -- in which the parity effect turns out to play a
much more important role. In this case the thermodynamic potential of 
the system $\Omega$ consists of two different contributions \cite{FN9}
\begin{equation}
\Omega =\Omega^{(r)}(\mu , T, \Phi_x,
\varphi)+\Omega^{(c)} (\mu , T, \varphi) \label{qpc}
\end{equation}
respectively from the bulk part of the ring and from QPC. The
optimal value of the phase difference $\varphi$ across QPC is
fixed by the condition $ \partial \Omega /\partial \varphi
=0$ which reads
\begin{equation}
-c\frac{\partial \Omega^{(r)}}{\partial
\Phi_x}=-\frac{2e}{\hbar} \frac{\partial
\Omega^{(c)}}{\partial \varphi}. \label{eqv}
\end{equation}
Here we made use of the fact that the thermodynamic potentials of
the ring depend both on $\Phi_x$ and $\varphi$ only via the
superfluid velocity $ v_{S}=(1/4\pi mR)(\varphi-2\pi
\Phi_x/\Phi_{0})$, in which case one can put $
\partial/\partial \Phi_x =-(2e/\hbar c)(\partial /\partial \varphi )
$. The left-hand side of Eq. (\ref{eqv}) represents the current
flowing inside the superconducting ring $I^{(r)}=-c\partial
\Omega^{(r)}\partial \Phi_x \simeq (ev_{F}N_{\rm
r}/L)(\varphi-2\pi \Phi_x /\Phi_{0})$. This value should be equal to
the current across QPC which is given by the right-hand side of
Eq. (\ref{eqv}). Estimating the maximum value of the latter for a
single channel QPC as $2e{\cal T}\Delta /\hbar$, we obtain
\begin{eqnarray}
\label{L<}\varphi&\simeq &2\pi \frac{\Phi_x}{\Phi_{0}},\;\;\;\mbox{  if  } \;\; L\ll L^{*},\\
\varphi&\simeq &2\pi n, \;\;\;\;\;\; \mbox{     if  }\;\; L\gg
L^{*}, \label{L>}
\end{eqnarray}
where $L^{*}=\xi_{0}N_{\rm r}/{\cal T} \gg \xi_0$. In a more
general case of QPC with $N$ conducting channels in the expression
for $L^{*}$ one should set ${\cal T} \to \sum_n^N{\cal T}_n$.

In what follows we will consider the most interesting limit $N \ll
N_{\rm r}$ and $L \ll L^{*}$. Due to Eq. (\ref{L<}) in this case
the dependence $I_{e/o}(\Phi_x)$ is fully determined by the
current-phase relation for QPS which can be found by means of Eq.
(\ref{Ie/o}) with $I_{f/b}=-(2e/\hbar)\partial
\Omega^{(c)}_{f/b}/\partial \varphi$. It is convenient to employ
the formula \cite{GZ}
\begin{equation}
I_{f/b}=\frac{2e}{\hbar}\sum_{n=1}^N T\sum_{\omega_{f/b}}
\frac{\sin\varphi}{\cos\varphi + W_n(\omega_{f/b})}. \label{fbfb}
\end{equation}
In the case of short QPS one has $W_n(\omega )= (2/{\cal
T}_n)(1+\omega^2/\Delta^2)-1$, where ${\cal T}_n$ is the transmission of
the $n$-th conducting channel. Substituting this function into
(\ref{fbfb}) and summing over $\omega_f$ one recovers the standard
result \cite{KO,Hab}
\begin{equation}
I_f(\varphi )=-\frac{2e}{\hbar}\sum_{n=1}^N\frac{\partial \varepsilon_n (\varphi)}
{\partial \varphi} \tanh\frac{\varepsilon_n (\varphi)}{2T},
\label{josephson}
\end{equation}
where
\begin{equation} \varepsilon_n (\varphi)=\Delta \sqrt{1-{\cal
T}_n\sin^2(\varphi/2)}. \label{Alev}
\end{equation}
The same summation over Bose Matsubara
frequencies $\omega_b$ yields
\begin{equation}
I_{b}=-\frac{2e}{\hbar}\sum_{n=1}^N \frac{\partial
\varepsilon_{n}(\varphi)} {\partial \varphi}
\coth\frac{\varepsilon_{n}(\varphi)}{2T}. \label{bose}
\end{equation}
Finally, the difference
$\Omega_b-\Omega_f\equiv \Omega_{bf}$ is evaluated as a sum of the
ring ($\Omega_{bf}^{(r)}$) and QPS ($\Omega_{bf}^{(c)}$)
contributions. The latter is found by integrating $I_{f/b}(\varphi
)$ over the phase difference $\varphi$:
\begin{equation}
\Omega_{bf}^{(c)}=2T\sum_{n=1}^N \ln
\coth\left(\frac{\varepsilon_{n}(\varphi)}{2T}\right),
\end{equation}
while the former is defined by the standard expression \cite{GZ94}
$$
\beta\Omega_{bf}^{(r)}=2{\cal V}\int \frac{d^3\mbox{\boldmath$p$}}
{(2\pi \hbar)^3}\ln\left(\coth
\frac{\varepsilon(\mbox{\boldmath$p$})}{2T} \right)\simeq \nu
{\cal V}\sqrt{\Delta T}e^{-\frac{\Delta }{T}}.
$$
Combining all these results with Eq. (\ref{Ie/o}) we get
$$
I_{e/o}=-\frac{2e}{\hbar}\sum_{n=1}^N \frac{\partial
\varepsilon_{n}(\varphi)} {\partial \varphi}
\tanh\frac{\varepsilon_{n}(\varphi)}{2T}
$$
\begin{equation}
\times \left[1\pm \frac{(\coth\frac
{\varepsilon_{n}(\varphi)}{2T})^{2}-1}{e^{\beta\Omega_{bf}^{(r)}}\prod\limits_{i=1}^N
(\coth\frac{\varepsilon_{i}(\varphi)}{2T}) ^{2} \pm 1}\right].
\label{result}
\end{equation}
Eq. (\ref{result}) represents the central result of this section. 
Together with 
Eq. (\ref{L<}) it establishes the complete dependence of PC on
the magnetic flux $\Phi$ in isolated superconducting nanorings with QPC.

Consider the most interesting limit $T \to 0$. In this case for the even
number of electrons in the ring PC is given by the expression 
(\ref{josephson}) which coincides identically with that for grand canonical 
ensembles \cite{KO,Hab}. On the other hand, for the odd number of electrons
PC will acquire an additional contribution which turns out to be
most important for the case of single channel QPS $N=1$. In that case
the expression in the square brackets of Eq. (\ref{result}) reduces to zero,
i.e. PC will be totally blocked by the odd electron. Thus, we predict {\it 
a novel mesoscopic effect} -- parity affected blocking of PC in superconducting
nanorings with QPC.

This result has a transparent physical
interpretation. Indeed, it is well known \cite{Furusaki} that the
result (\ref{josephson}) can be expressed via the contributions
of discrete Andreev levels $E_{\pm}(\varphi )=\pm \Delta {\cal
D}(\varphi )$ inside QPS as
\begin{equation}
I(\varphi )=\frac{2e}{\hbar}\left[\frac{\partial E_-}{\partial
\varphi} f_-(E_-) +\frac{\partial E_+}{\partial \varphi}
f_+(E_+)\right], \label{jos2}
\end{equation}
where ${\cal D}(\varphi )$ is defined in Eq. (\ref{D1}).
Using the Fermi filling factors for these levels $f_\pm (E_\pm
)=[1+\exp (E_{\pm} (\varphi )/T)]^{-1}$ one arrives at Eq.
(\ref{josephson}). If we now fix the number of electrons inside the ring and
consider the limit $T \to 0$ the filling factors will be modified as follows. 
For the even number ${\cal N}$ all
electrons are paired occupying states with energies below the
Fermi level. In this case one has $f_- (E_- )=1$, $f_+ (E_+ )=0$, 
the current is entirely determined by the
contribution of the quasiparticle state $E_-$ and Eq. (\ref{jos2})
yields the same result as one for the grand canonical ensemble. By
contrast, in the case of odd number of electrons one electron
always remains unpaired and occupies the lowest available energy
state -- in our case $E_+$ -- above the Fermi level. Hence, for
odd ${\cal N}$ one has $f_\pm (E_\pm )=1$, the
contributions of two quasiparticle energy states in Eq.
(\ref{jos2}) exactly cancel each other, and the current across QPS
remains zero for any $\varphi$ or the magnetic flux $\Phi_x$. This is just the
blocking effect which we have already obtained above from a more formal
consideration.

For $N >1$ and/or at non-zero temperatures this parity-affected blocking of 
PC becomes incomplete. But also in this case the parity effect remains
essential at temperatures $T <T^*$ substantially affecting, e.g., the
current-phase relation for QPC. For $T >0$ this relation will deviate from
the grand canonical one both for even and odd ensembles \cite{SZ}.  

Finally, we turn to superconducting rings
containing a piece of a normal metal. Here we only restrict our
attention to transparent $SNS$ junctions with length of the normal
metal $d \gg \xi_0$. In this case for $\omega \ll \Delta$ one has
$W_i(\omega )= \cosh (2\omega d/v_F)$. Substituting this function
into (\ref{fbfb}) and repeating the the whole calculation as
above, in the limit $T \to 0$  we obtain
\begin{equation}
I_e=\frac{ev_FN}{\pi d}\varphi
,\;\;\;\;I_o=\frac{ev_FN}{\pi d}\left(\varphi -\frac{\pi
\mbox{sgn} \varphi}{N}\right). \label{SNS}
\end{equation}
These results apply for $-\pi <\varphi<\pi$ and should be
$2\pi$-periodically continued otherwise. We observe that the
current $I_e$ again coincides with that for the grand canonical
ensembles \cite{Kulik}, while for odd ${\cal N}$ the current-phase
relation is shifted by the value $\pi /N$. This shift has a simple
interpretation as being related to the odd electron contribution
$(2e/\hbar )\partial E_0/\partial \varphi$ from the lowest (above
the Fermi level) Andreev state $E_0(\varphi )$ inside the $SNS$ junction.
Unlike in QPC, this contribution does not compensate for the
current from other quasiparticle states. Rather it provides a
possibility for a parity-induced $\pi$-junction state in our
system. Indeed, according to Eq. (\ref{SNS}) for single mode $SNS$
junctions the ``saw tooth'' current-phase relation will be shifted
exactly by $\pi$. For more than one conducting channel $N>1$
within the interval $-\pi <\varphi <\pi$ there exists a twofold
degenerate minimum energy (zero current) state \cite{BGZ}
occurring at $\varphi =\pm \pi /N$. In the particular case $N=2$
the current-phase relation $I_o(\varphi )$ turns $\pi$-periodic.

The well known feature of superconducting rings interrupted by a 
$\pi$-junction
is the possibility to develop {\it spontaneous} supercurrent in the 
ground state
\cite{Leva}. Although this feature is inherent to any type of $\pi$-junctions, 
in the case of the standard sinusoidal current-phase relation such
spontaneous supercurrents can occur only for sufficiently large values
of the ring inductance ${\cal L}$ \cite{Leva}. In contrast, in the 
situation studied here
the spontaneous
current state is realized for {\it any} inductance of the ring
because of the non-sinusoidal dependence $I_o(\varphi )$
(\ref{SNS}).

In order to demonstrate that let us assume that no external flux
is applied to our system. Then at $T \to 0$ the energy of an $SNS$ ring with
odd number of electrons can be written in the form
\begin{equation}
 E_o=\frac{\Phi^2}{2c{\cal L}}+
\frac{\pi\hbar v_FN}{\Phi_0^2d}
\left(\Phi -\frac{\Phi_0\mbox{sgn}\Phi}{2N}\right)^2,
\label{ESNS}
\end{equation}
where $\Phi$ is the flux related to the circular current flowing in the ring.
Minimizing this energy with respect to $\Phi$, one easily observes that
a non-zero spontaneous current   
\begin{equation}
I=\pm \frac{e v_F}{d}\;\left[1+
\frac{2ev_FN}{d}\frac{{\cal L}}{\Phi_0}\right]^{-1}
\label{spcur}
\end{equation}
should flow in the ground state of our system. This is yet one more
remarkable consequence of the parity effect: Just by changing
${\cal N}$ from even to odd one can induce non-zero PC without any
external flux $\Phi_x$. In the limit of small inductances
${\cal L} \ll \Phi_0d/ev_FN$ -- which is easy to reach in the
systems under consideration -- the value of $I$ does not depend
on the number of channels $N$ and is given by the universal
formula $I=\pm ev_F/d$. For generic parameters this value can easily be
as large as $I \sim 10$ na.

In summary, new physical effects emerge from an interplay between
the electron parity number and persistent currents in
superconducting nanorings. These effects can be directly tested in
modern experiments and possibly used for engineering new types of
superconducting flux-charge qubits.

\section{Acknowledgments}

The author gratefully acknowledges collaboration and numerous discussions with
A.V. Galaktionov, D.S. Golubev and S.V. Sharov. This work is part of the 
Kompetenznetz ``Funktionelle Nanostructuren'' supported by the Landestiftung 
Baden-W\"urttemberg gGmbH and of the STReP ``Ultra-1D'' supported by 
the EU.

\end{multicols}

\begin{references}
\bibitem{FN} This deviation is not as strong as in the ballistic case:
The supercurrent in long diffusive junctions reaches its maximum at the value
$\varphi \simeq 1.27 \pi /2$ \cite{us}.
\bibitem{FN9} The relation (\ref{qpc}) is strictly applicable only for 
grand canonical
ensembles. However, at this point the difference between 
canonical and grand canonical ensembles is unimportant and can be disregarded.
\end{references}
\end{document}